\documentclass[sort&compress
    ,final            
 ]
  {aipproc}

\layoutstyle{6x9}

\newcommand{\be}{\begin{equation}}
\newcommand{\ee}{\end{equation}}
\newcommand{\bea}{\begin{eqnarray}}
\newcommand{\eea}{\end{eqnarray}}
\newcommand{\open}{{<\kern -0.3 em{\scriptscriptstyle )}}}

\begin{document}

\title{Transversity: theory and phenomenology\footnote{Invited talk delivered at the ``International Workshop on Diffraction in High-Energy Physics'' (DIFFRACTION~2012), Sept.~10-15, 2012, Puerto del Carmen, Lanzarote, Canary Islands (Spain).}}


\author{Umberto D'Alesio}{
  address={Dipartimento di Fisica, Universit\`a di Cagliari, Cittadella Universitaria, and
  Istituto Nazionale di Fisica Nucleare, Sezione di Cagliari, C.~P.~170, I-09042 Monserrato (CA), Italy}
}

\begin{abstract}
The distribution of transversely polarized quarks inside a transversely polarized nucleon, known as \emph{transversity}, encodes a basic piece of information on the nucleon structure, sharing the same status with the more familiar unpolarized and helicity distributions. I will review its properties and discuss different ways to access it, with highlights and limitations. Recent phenomenological extractions and perspectives are also presented.
\end{abstract}

\maketitle


\section{Theory}

Three distributions characterize, in a collinear framework, the internal structure of a fast moving nucleon: the momentum ($q(x)$), the helicity ($\Delta q(x)$) and the transversity ($\Delta _Tq(x)$ or $h_1^q(x)$)~\cite{Barone:2001sp} distributions. On equal footing on the theory side, the first two have been under active experimental investigation for decades. By contrast, $\Delta_Tq$, while studied in a large class of models, only recently has been accessed phenomenologically.

In a partonic picture, given a transversely polarized nucleon (w.r.t.~its direction of motion) and denoting with $q_{\uparrow\downarrow}$ the number density of quarks with polarization parallel (antiparallel) to that of the nucleon, the transversity is the difference $q_\uparrow-q_\downarrow$. More formally, $\Delta_Tq(x)$ is given in terms of a hadronic matrix element of a nonlocal operator ($\langle \bar\psi(0)i\sigma^{1+}\gamma_5 \psi(\xi^-) \rangle$) and is a leading-twist quantity, like $q(x)$ and $\Delta q(x)$. If expressed in the helicity basis, where $q(x)$ and $\Delta q(x)$ are difference of probabilities, transversity appears as the interference of off-diagonal amplitudes, showing its chiral-odd nature.\footnote{Notice that this emerges also from its operator structure.} This explains the difficulties in measuring it, being, in fact, unaccessible via the helicity (chirality) conserving inclusive deep inelastic scattering (DIS) processes. To measure $\Delta_Tq$, the chirality must be flipped twice, hence another chiral-odd partner is required.

It is the only source of information on the tensor charge ($\int (\Delta_T q - \Delta_T \bar q)$), a fundamental charge as important as the vector ($\int (q - \bar q)$) and the axial ($\int (\Delta q + \Delta \bar q)$) charges. It obeys a nontrivial bound~\cite{Soffer:1994ww}, $|\Delta_T q|\le (q +\Delta q$)/2, and does not couple to gluons, implying a non-singlet $Q^2$-evolution (and its strong suppression at low $x$); moreover, no gluon transversity exists for a spin-1/2 hadron.

\section{Probing transversity}

Its chiral-odd nature demands for processes with at least two hadrons.
Two types of observables can be considered: 1) a double transverse spin asymmetry (DtSA) - the simplest case from the theory side - where the chiral-odd partner is a second transversely polarized hadron, either in the initial (a) or in the final (b) state; 2) a single spin asymmetry (SSA), that can be analyzed in two frameworks: (a) one based on transverse momentum dependent distributions (TMDs), encoding correlations between spin and intrinsic transverse momenta ($k_\perp$); (b) the other based on dihadron fragmentation functions (DiFFs), encoding the interference between different partial waves of a dihadron system. We will not discuss other possible ways to access it, like through higher-twist functions or higher-spin particles ($\rho$ mesons). Notice that only the option (1a) represents a self-sufficient observable, involving only the transversity distribution (``squared''). The other cases require, unavoidably, the appearance of extra unknown soft functions.

\subsection {Double transverse spin asymmetries}
Within the first class of observables a major role is played by the Drell-Yan (DY) process with transversely polarized protons, $p^\uparrow p^\uparrow\to l^+l^-\,X$, as proposed in Ref.~\cite{Ralston:1979ys} more than thirty years ago. The double spin asymmetry is given as
\be
A_{TT} \equiv \frac{d\sigma^{\uparrow\uparrow} -
d\sigma^{\uparrow\downarrow}} {d\sigma^{\uparrow\uparrow} +
d\sigma^{\uparrow\downarrow}} \sim \sum_q e_q^2 \left[ h_1^q(x_1) \,
h_1^{\bar q}(x_2) +  h_1^{\bar q}(x_1) \, h_1^q(x_2)\right]\,.
\ee
However, for RHIC kinematics, due to the the low-$x$ region probed together with the expected small values of $h_1^q$ for antiquarks, one gets an upper bound of only 1-2\%~\cite{Martin:1999mg}. Much larger asymmetries can be obtained by using transversely polarized antiprotons ($p^\uparrow \bar p^\uparrow$), as proposed by the PAX Collaboration~\cite{Barone:2005pu}, with $A_{TT}$ given as a product of two quark transversity distributions in the valence region. Here the problem is the too low counting rates expected, possibly circumvented by looking at the $J/\psi$ peak~\cite{Anselmino:2004ki}, with a gain of a factor two in statistics. Other DtSAs, like those for inclusive photon, jet or pion production, are strongly suppressed by the gluon dominance in their denominators.

For processes with a final polarized hadron, $\Lambda$ production in semi-inclusive deep inelastic scattering (SIDIS), or in $pp$ collisions, plays a special role, thanks to the self-analyzing power of $\Lambda$. The price is the unknown transversely polarized fragmentation function $H_1^q$ coupling to $\Delta_T q$. Moreover, the dominance of $u$ quarks in the proton and of strange quarks in the fragmentation mechanism might implies low values for the spin transfer $D_{NN}$. Complementary information on $H_1^q$ can be obtained from the study of $e^+e^-\to\Lambda^\uparrow \bar \Lambda^\uparrow X$ as suggested in~\cite{Contogouris:1995xc}.

\subsection{Single spin asymmetries}

$\bullet$ {$\Delta_T q$ coupled to TMDs}\\
When the intrinsic transverse momentum is not integrated out, at leading twist there are eight TMDs for a nucleon (three of which, upon integration over $k_\perp$, yield $q(x)$, $\Delta q(x)$ and $\Delta_T q(x)$).
In DY and SIDIS processes, for which TMD factorization has been proved to hold~\cite{Ji:2004wu,Ji:2004xq}, by looking at specific azimuthal dependences, one is able to disentangle unambiguously those contributions involving $\Delta_T q$. More precisely, in DY one can access the transversity coupled with the Boer-Mulders function~\cite{Boer:1999mm}. This unknown  distribution gives the probability to find a transversely polarized quark inside an unpolarized proton and can be extracted from the $\cos2\phi$ dependence in the unpolarized DY cross section.
In SIDIS, with a transversely (T) polarized target, a peculiar modulation in the azimuthal dependence of the final hadron involving $\Delta_Tq$ emerges, the so-called Collins effect~\cite{Collins:1992kk, Mulders:1995dh, Bacchetta:2006tn,Anselmino:2011ch}:
\be
A_{UT}\sim \cdots + \Delta_Tq\otimes H_1^{\perp q} \sin (\phi_h+\phi_S)\,,
\ee
where $\phi_S$, $\phi_h$ are the azimuthal angles, respectively, of the proton spin and the final hadron momentum. The extra unknown, $H_1^{\perp q}$, is the Collins function~\cite{Collins:1992kk}, giving the probability for a transversely polarized quark to fragment into an unpolarized hadron. It can be extracted from the azimuthal asymmetries in the distribution of two almost back-to-back hadrons in $e^+e^-$ annihilation~\cite{Boer:1997mf}. The experimental evidence of these asymmetries~\cite{Airapetian:2004tw,Abe:2005zx} has indeed allowed for the first-ever extraction of $\Delta_Tq$~\cite{Anselmino:2007fs}.

Another possible source of information on transversity, within a TMD factorization scheme (even if still not formally proven), is the study of the azimuthal distribution of a pion inside a jet in $p^\uparrow p$ collisions~\cite{Yuan:2007nd, D'Alesio:2010am}. At variance with the single-inclusive particle production, where all effects can mix up together, here, by taking suitable moments of the asymmetry, one can single out the Collins effect and probe $\Delta_Tq$.

Before completing this section a word of caution on the analysis of SSAs within the TMD approach is mandatory. Indeed, beyond the tree-level approximation, TMD factorization involves an extra soft factor, that, depending on the scale, implies a dilution of the asymmetry at large $Q^2$~\cite{Boer:2008fr}. Some developments on TMD evolution have recently appeared~\cite{Collins:2011zzd,Aybat:2011ge,Anselmino:2012aa,GarciaEchevarria:2011rb}, but its quantitative effect (in particular for chiral-odd functions, like $H_1^{\perp q}$) is still under investigation and not taken into account in phenomenological extractions.\\

$\bullet$ {$\Delta_Tq$ coupled to DiFFs}\\
The quark fragmentation mechanism $q^\uparrow\to \pi\pi$, e.g.~in the process $lp^\uparrow\to l' (\pi\pi) \, X$, can act as a polarimeter, allowing for the extraction of $h_1^q$ in combination with the chiral-odd dihadron fragmentation function $H_1^\open$~\cite{Jaffe:1997hf, Radici:2001na}. This extra unknown, describing the correlation between the transverse polarization of the fragmenting quark and the azimuthal orientation of the plane containing the momenta of the detected hadron pair, can be extracted by analyzing, for instance, double azimuthal correlations in $e^+e^-\to (\pi\pi)_1 (\pi\pi)_2\, X$~\cite{Artru:1995zu}. Notice that this method relies on the standard collinear factorization with proper QCD evolution of DiFFs.

\begin{figure}[t!]
\centering
\mbox{}\hspace*{-4mm}\begin{minipage}[c]{.34\textwidth}
\centering
\includegraphics[width=1\textwidth, angle=-90]{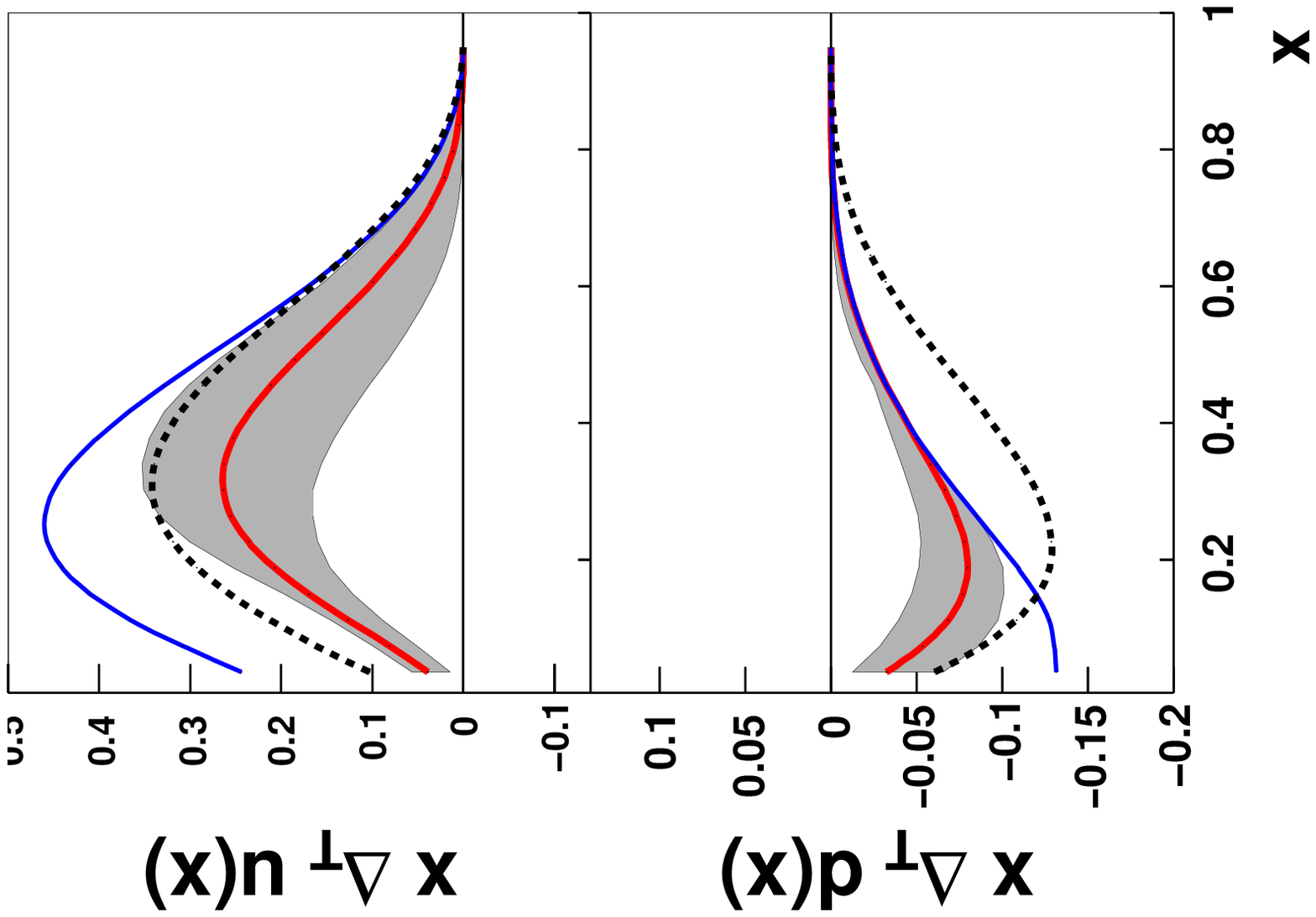}
\end{minipage}%
\hspace{-1mm}%
\begin{minipage}[c]{.34\textwidth}
\centering
\includegraphics[width=1\textwidth]{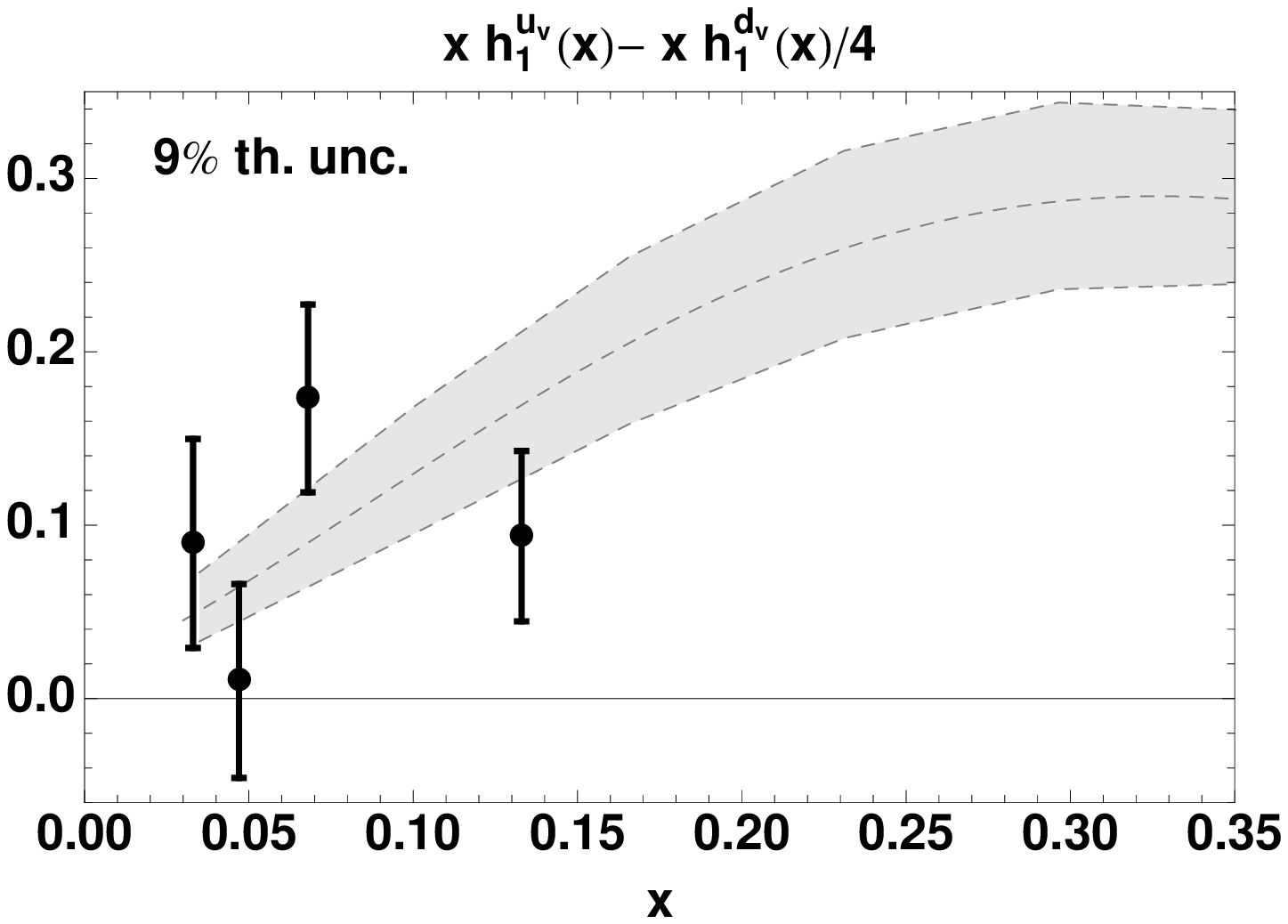}
\end{minipage}
\begin{minipage}[c]{.34\textwidth}
\centering
\includegraphics[width=1\textwidth]{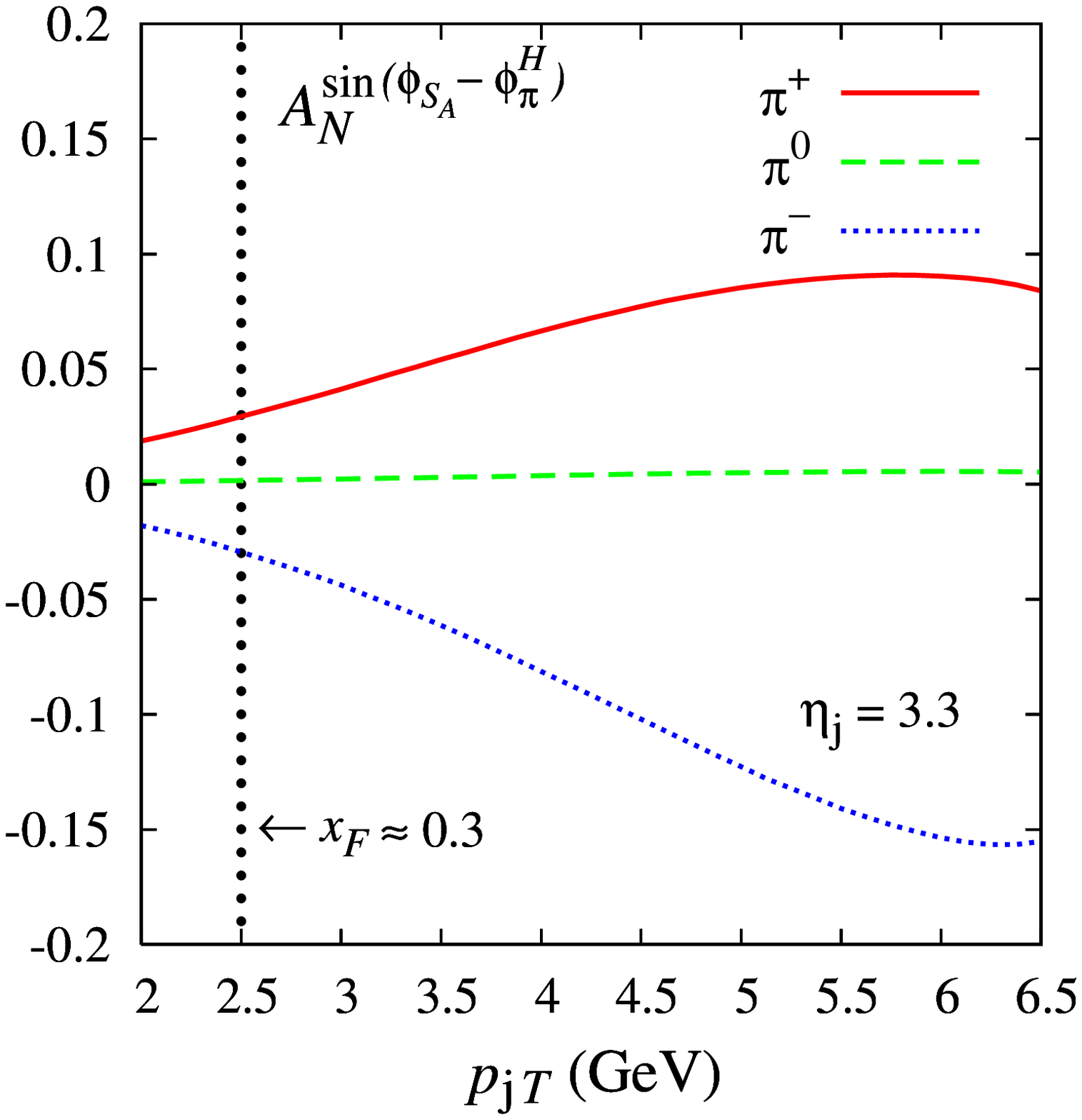}
\end{minipage}
\caption{Left panel: $\Delta_Tq(x)$ (solid red line) from the fit~\cite{Anselmino:2008jk} on the Collins effect. Helicity distribution (dotted black line), Soffer bound (solid blue line) and uncertainty band are also shown. Central panel: Results from Ref.~\cite{Bacchetta:2011ip} together with the uncertainty band as deduced from the fit in the left panel. Right panel: Predictions for the Collins effect in $A_N$ in $pp\to {\rm jet}\, \pi\, X$~\cite{D'Alesio:2010am} based on the TMD fit.
}
\end{figure}

\section{Present status and perspectives}

The impressive experimental activity, in the last decade, of various collaborations, HERMES~\cite{Airapetian:2004tw, Airapetian:2010ds} and \cite{Airapetian:2008sk}, COMPASS~\cite{Alexakhin:2005iw, Alekseev:2010rw}, and Belle~\cite{Abe:2005zx,Seidl:2008xc} and \cite{Vossen:2011fk}, has allowed, at last, for two independent phenomenological extractions of $\Delta_T q$.

The main aspects of the analysis of the Collins effect~\cite{Anselmino:2007fs, Anselmino:2008jk}, can be summarized as follows: $i)$ a power-like ($x^a(1-x)^b$) parametrization of $\Delta_Tq$ for up and down quarks and, similarly, of the favored ($u\to\pi^+$) and unfavored ($d\to\pi^+$) Collins fragmentation functions; $ii$) a factorized gaussian $k_\perp$ dependence; $iii)$ universality of $H_1^{\perp q}$~\cite{Metz:2002iz, Collins:2004nx}; $iv)$ simultaneous extraction of $\Delta_Tq$ and $H_1^\perp$ from a global fit of SIDIS and $e^+e^-$ data; $v)$~proper $Q^2$ evolution for $\Delta_Tq$, while replaced with that of the unpolarized fragmentation function for $H_1^{\perp q}$. The results of the fit~\cite{Anselmino:2008jk}, at $Q^2=2.4$ GeV$^2$, are shown in Fig.~1 (left panel), where one can see how $\Delta_T q$ (solid red line) is sizeable, different from $\Delta q$ (dotted black line) and much smaller than the Soffer bound~\cite{Soffer:1994ww} (solid blue line).

On the complementary side, i.e.~the study via DiFFs~\cite{Bacchetta:2011ip}, the main features are:
$i)$ $H_1^{\open\, u} = - H_1^{\open\, d}$ for charged pion pairs, based on isospin symmetry and charge conjugation; $ii)$ unpolarized $D^{q\to\pi\pi}$ from PITHYA, due to the absence of data;
$iii)$ extraction of $H_1^\open$ for $(\pi^+\pi^-)$ from Belle $e^+e^-$ data;  $iv)$ proper $Q^2$-evolution of $H_1^\open$ from 110 to 2.4 GeV$^2$; $v)$~extraction of the combination $(xh_1^u-xh_1^d/4)$ from SIDIS data.
The results of this fit, with their statistical errors, are presented in Fig.~1 (central panel), where the uncertainty band of the fit via TMDs is also shown. The compatibility of the two extractions, taking into account the various assumptions behind them, is quite encouraging.

Finally, in the right panel of Fig.~1 we show some predictions for the Collins effect in the azimuthal distribution of pions inside a jet in $pp$ collisions at $\sqrt s = 200$ GeV and large rapidities~\cite{D'Alesio:2010am}. This process, under active investigation at RHIC, could play a double role: testing the universality of the Collins function in $pp$~collisions and extending the coverage of $\Delta_Tq$ to the large $x$ region where it is still poorly known.

Summarizing, thanks also to recent phenomenological analysis, transversity has definitely reached its full status in high-energy spin physics. DtSAs, difficult experimentally, remain still the cleanest observable on the theory side, with $A_{TT}$ in DY processes being the golden channel.
Concerning SSAs within a TMD approach, improvements in the proper $Q^2$-evolution are mandatory in view of global fits, whilst the complementary extraction through DiFFs requires further work. More and more precise data (e.g.~at JLab in the still unexplored large $x$ region) are eagerly awaited.


\begin{theacknowledgments}
The author thanks the organizers for their kind invitation to such a nice workshop.
\end{theacknowledgments}



\bibliographystyle{aipproc}   


\hyphenation{Post-Script Sprin-ger}

\end{document}